\documentclass[journal]{IEEEtran}
\usepackage{cite}

\ifCLASSINFOpdf
  \usepackage[pdftex]{graphicx}
\else
  \usepackage[dvips]{graphicx}
\fi
\usepackage{multirow}

\usepackage{amsmath}
\begin{document}
%
\title{Low Precision Floating-point Arithmetic for High Performance FPGA-based CNN Acceleration}
%
%
%

\author{Chen~Wu, 
		Mingyu~Wang, 
		Xinyuan~Chu, 
		Kun~Wang, 
		Lei~He
\thanks{C. Wu, M. Wang, X. Chu, K. Wang and L. He are with the Department of Electrical and Computer Engineering, University of California, Los Angeles, CA, 90095,  
		USA (e-mail: chenwu1989@ucla.edu, mingyuw@ucla.edu, xinyuansjtu@gmail.com, wangk@ucla.edu, lhe@ee.ucla.edu).}}

\maketitle

\begin{abstract}
Low precision data representation is important to reduce storage size and memory access for convolutional neural networks (CNNs). Yet, existing methods have two major limitations: (1) requiring re-training to maintain accuracy for {\it deep} CNNs, and (2) needing 16-bit floating-point or 8-bit fixed-point for a good accuracy.
In this paper, we propose a low precision (8-bit) floating-point (LPFP) quantization method for FPGA-based acceleration to overcome the above limitations. Without any re-training, LPFP finds an optimal 8-bit data representation with negligible top-1/top-5 accuracy loss (within 0.5\%/0.3\% in our experiments, respectively, and significantly better than existing methods for {\it deep} CNNs). Furthermore, we implement one 8-bit LPFP multiplication by one 4-bit multiply-adder (MAC) and one 3-bit adder, and therefore implement {\it four} 8-bit LPFP multiplications using one DSP slice of Xilinx Kintex 7 family (KC705 in this paper) while one DSP can implement only {\it two} 8-bit fixed-point multiplications. Experiments on six typical CNNs for inference show that on average, we improve throughput by $64.5\times$ over Intel i9 CPU and by $1.5\times$ over existing FPGA accelerators. Particularly for VGG16 and YOLO, compared to six recent FPGA accelerators, we improve average throughput by 3.5$\times$ and 27.5$\times$ and improve average throughput per DSP by 4.1$\times$ and 5$\times$, respectively. To the best of our knowledge, this is the first in-depth study to simplify one multiplication for CNN inference to one 4-bit MAC and implement four multiplications within one DSP while maintaining comparable accuracy without any re-training.
\end{abstract}


%
\IEEEpeerreviewmaketitle

\section{Introduction}
\label{section:introduction}
\IEEEPARstart{C}{onvolutional} neural networks (CNNs) have demonstrated a breakthrough in performance for a broad range of applications including object recognition \cite{obj_recog}, object detection \cite{obj_detec} and speech recognition \cite{spe_recog}. However, CNNs often have huge computation complexity. This motivates accelerating CNNs by CPU/GPU clusters \cite{cluster}, FPGAs \cite{fpga} and ASICs \cite{diannao}. Customized accelerators/processors on FPGAs have shown more promising throughput and power efficiency than traditional CPU/GPU clusters \cite{caffeine, systolic}.

\begin{figure}[t]
\centering
\includegraphics[width=3.5in]{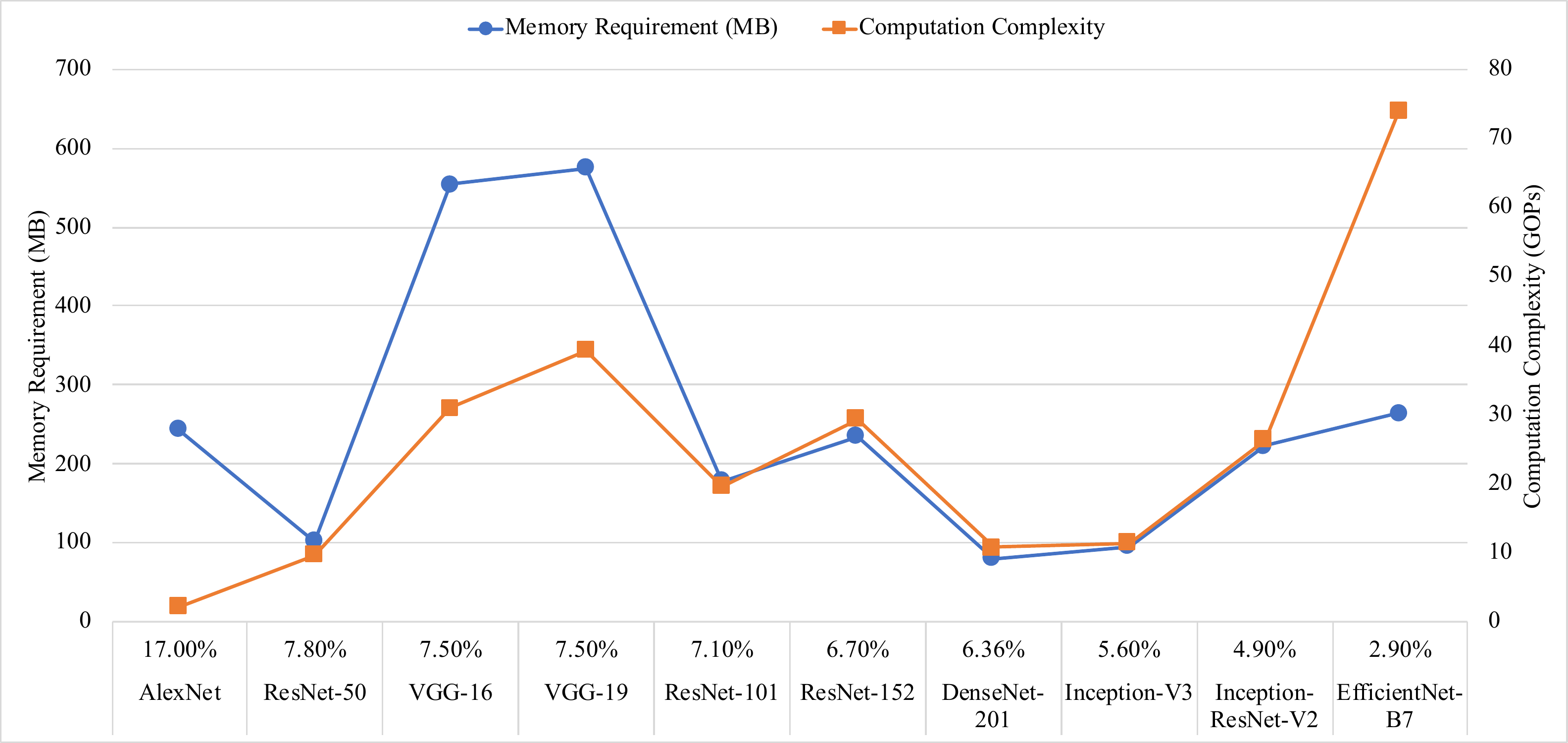}
\caption{Computation complexity and memory requirement with respect to different CNNs.}
\label{fig:cc_param}
\end{figure}

Larger and deeper CNNs have been developed to improve performance for a broader range of scenarios. For example, the top-5 error for ImageNet \cite{imagenet} classification decreases from 17\% to 2.9\%. However, computation complexity and number of parameters increase dramatically as depicted in Figure \ref{fig:cc_param}. To be specific, the computation complexity of a feed-forward process of a 224$\times$224 RGB image increases from 2.27 GOP of AlexNet \cite{alexnet} in 2012 to 74 GOP of EfficientNet-B7 \cite{efficientnet} in 2019. At the same time, the number of parameters stays large at 264 MB. Such great computation complexity makes it harder for general-purpose processor to meet the requirements of real-time applications. On the other hand, the great quantities of parameters lead to a big challenge for communication between off-chip and on-chip memories because of bandwidth constraints. 

There are two types of research to reduce computation and parameter complexities for CNN inference. The first one is deep compression including weight pruning, weight quantization and compression storage \cite{han_pruning, han_compression}. However, irregularity caused by deep compression degrades parallelism and hardware performance. Cambricon-S \cite{cambricon_s} alleviates irregularity in sparse neural networks through a software/hardware co-design approach to improve hardware performance. However, all the above accelerators need time-consuming re-training process to maintain accuracy.

\begin{table}
  \centering
  \caption{Resource utilization of multipliers on FPGA for different data representations. DSP: digital signal processing, LUT: look-up table, FF: flip-flop. $M4E3$: 1-bit sign, 4-bit mantissa and 3-bit exponent.}
  \begin{tabular}{c|c|c|c}  \hline
    Data Representation                                          &DSP      &LUT     &FF    \\ \hline
    {\bf one} 16-bit floating multiplication                     &1        &85      &167   \\ \hline
    {\bf one} 16-bit fixed multiplication                        &1        &0       &0     \\ \hline
    {\bf two} 8-bit fixed multiplications                        &1        &2       &0     \\ \hline
    {\bf four} 8-bit floating ($M4E3$) multiplications           &1        &20      &27    \\ \hline
  \end{tabular}
  \label{table:mul_dsp}
\end{table}

The second type of research is more efficient data representation, also known as quantization for circuit implementation. \cite{16_fp} used 16-bit floating-point in contrast to 32-bit commonly used for computing. However, one 16-bit floating-point multiplier on FPGA needs 1 DSP, 85 LUTs and 167 FFs when using Xilinx floating-point IP \cite{xilinx_ip} as shown in Table~\ref{table:mul_dsp}, leading to a low hardware efficiency. Since one 16-bit or smaller fixed-point multiplier can be fit into one DSP, both 16-bit \cite{16_fix_1, 16_fix_2} and 8-bit \cite{roofline, tpu, diannao, opu} fixed-point were employed to gain more hardware efficiency than 16-bit floating-point does. Another 8-bit arithmetic, called block floating-point (BFP), was also applied \cite{bfp_1, bfp_fpga}, where a parameter has its own mantissa but shares a same exponent for one data block. \cite{ARM_mixed} proposed a mixed data representation with floating-point for weights and fixed-point for activations ({\it e.g.,} outputs of a layer). \cite{xilinx_float} developed an 8-bit floating-point quantization scheme, which needs an extra inference batch to compensate for the quantization error. However, \cite{ARM_mixed} and \cite{xilinx_float} did not present a circuit design for their approaches. While all aforementioned work has a good accuracy with re-training, more aggressive data representations such as binary \cite{bin1}, ternary \cite{ter1}, and mixed precision (2-bit activations and ternary weights) \cite{mixed} may suffer from great accuracy loss even with time-consuming re-training.

In this paper, we first propose a low precision floating-point (LPFP) to quantize both weights and activations. During the quantization process, an optimal LPFP data format and the corresponding scale factor are decided for a workload of CNNs. Our proposed quantizer works for {\it deep} CNNs (more than 100 {\it convolutional/fully-connected} layers). On average, the top-1 accuracy loss is within 0.5\%, while V-Quant \cite{V-Quant} that works for such {\it deep} CNNs has a top-1 accuracy loss about 1\% with fine-tuning. Then, we design a LPFP based FPGA processor to further improve the performance for CNN inference. We are able to implement four 8-bit floating-point multiplications within one DSP (see Table~\ref{table:mul_dsp}). We experimented for inference of AlexNet, VGG16 \cite{vgg}, ResNet50/101/152 \cite{resnet_v1} and DenseNet201 \cite{densenet} via Xilinx KC705. We can achieve an average throughput of 1100.4 GOPS (Giga-Operations Per Second), and it is 1.43 GOPS per DSP. Moreover, the average throughput for these networks is 64.5$\times$ and $1.5\times$ over Intel i9 CPU and existing accelerators, respectively. Compared with six existing accelerators for VGG16 and YOLO, on average, our processor improves throughput by 3.5$\times$ and 27.5$\times$, while improveing per DSP throughput by 4.1$\times$ and 5$\times$, respectively. 

While existing work needs re-training (including calibration \cite{xilinx_float} and fine-tuning \cite{rna}), our quantization method does not need any re-training to compensate for quantization error. Furthermore, to the best of our knowledge, this is the first work that can fit four 8-bit multiplications for inference in one DSP while maintaining comparable accuracy without any re-training.

\section{Background and Motivation}
\label{section:background and motivation}

\subsection{Background}
\label{subsection:background}

\subsubsection{CNNs}
\label{subsubsection:cnns}

CNNs are used to classify or recognize objects by passing the inputs through multiple types of layers. In each layer, multiple neurons are constructed to process different inputs and pass the outputs to the next layer through connections, and the connections are used to store the weights for the network. Based on different processing procedures, the layers are typically divided into {\it convolutional, pooling, activation, normalization, fully-connected, residual} and {\it inception} layers. Among them, {\it convolutional/fully-connected} layers consume most portions of computation while {\it fully-connected} layers require largest memory to store weights. According to this, we divide the size of CNNs into three categories with respect to the number of {\it convolutional/fully-connected} layers: 1) {\it slim} for less than 50 layers, 2) {\it medium} for 50 to 100 layers, and 3) {\it deep} for more than 100 layers, as shown in Table~\ref{table:benchmark} where we report the detailed network information.

\begin{table}
  \centering
  \caption{Characteristics of CNN benchmarks. GOP is giga-operations needed by one 224$\times$224 RGB image.}
  \begin{tabular}{c|c|c|c}  \hline
    {\bf CNN}              &{\bf Type}         &{\bf Operations}   &{\bf Model Weights}   \\ \hline
    AlexNet                &{\it slim}         &2.27 GOP           &249.51 MB             \\ \hline
    VGG16                  &{\it slim}         &30.94 GOP          &553.43 MB             \\ \hline
    ResNet50               &{\it medium}       &9.74  GOP          &46.05 MB              \\ \hline
    ResNet101              &{\it medium}       &19.70 GOP          &166.37 MB             \\ \hline
    ResNet152              &{\it deep}         &29.39 GOP          &229.39 MB             \\ \hline
    DenseNet201            &{\it deep}         &10.85 GOP          &68.63 MB              \\ \hline
  \end{tabular}
  \label{table:benchmark}
\end{table}

\subsubsection{Low Precision Floating-point}
\label{subsubsection:LFP}

Similar to the definition of 32-bit floating-point from the IEEE-754 standard \cite{IEEE754}, the binary representation of LPFP number comprises {\it sign, mantissa} and {\it exponent} in order. The decimal value of LPFP number is then calculated by:
\begin{equation}
\label{equation:normal}
    V_{dec} = (-1)^S\times1.M\times2^{E-E_b},
\end{equation}
where $V_{dec}$ is the value in decimal, $S, M$ and $E$ are all unsigned values and denote the {\it sign, mantissa} and {\it exponent}, respectively. For exponent bias $E_b$ in Eq. ~(\ref{equation:normal}), it is introduced to both positive and negative exponents as
\begin{equation}
\label{equation:bias}
    E_b = 2^{DW_E-1}-1,
\end{equation}
where $DW_E$ is the data width of $E$. Different from the IEEE Standard, data widths for $M$ and $E$ in this paper are not fixed. In later sections, we use the term $MaEb$ to indicate different combinations, where $a$ and $b$ indicate the bit width of $M$ and $E$, respectively. For example, $M3E4$ means the mantissa is 3 bits while the exponent is 4 bits. 

There are three special definitions in IEEE-754 standard. The first is subnormal numbers when $E=0$, then Eq. ~(\ref{equation:normal}) is modified to:
\begin{equation}
\label{equation:subnormal}
    V_{dec} = (-1)^S\times0.M\times2^{1-E_b}.
\end{equation}
Note that Infinity (Inf) and Not a Number (NaN) are the other two special cases, but are not used in our work. This is because our saturation scheme saturates large numbers to the maximal number, as illustrated in detail in Subsection~\ref{subsection:quantization_process}.

\subsection{Motivation}
\label{subsection:motivation}

CNN accelerators with lower data width have significant improvements in terms of memory size, memory bandwidth and power efficiency. Due to the lack of floating-point arithmetic units in FPGA, researchers have used low precision fixed-point instead of floating-point. A 16-bit fixed-point quantization to find the best scale factor for each layer was proposed in \cite{16_fix_1}. However, this required time-consuming re-training to amend the weights to maintain accuracy. Furthermore, a model was developed to quantitatively analyze the convolution loops and optimize design objectives such as memory access and latency \cite{16_fix_2}. However, it had an accuracy loss as large as 2\%. A shared drawback for the above two approaches is the low per DSP throughput (0.279 GOPS/DSP for \cite{16_fix_1} and 0.472 GOPS/DSP for \cite{16_fix_2}) because of using 16-bit multiplication.

An 8-bit fixed-point accelerator was designed in \cite{angle_eye} for embedded FPGAs, with a low per DSP throughput of 0.444 GOPS/DSP. DNNBuilder \cite{dnnbuilder} aimed to automatically build high-performance DNN hardware accelerators for both cloud- and edge-FPGAs with 8-bit fixed-point quantization. It increased the per DSP throughput to 0.771 GOPS by better architecture exploration; however, its quantization method incurred 4.6\% top-1 accuracy degradation without fine-tuning. FPGA accelerator with the aforementioned BFP arithmetic \cite{bfp_fpga} had a per DSP throughput of 0.741 GOPS. However, only {\it slim} and {\it medium} CNNs were validated in their approach. In short, existing approaches cannot improve the per DSP throughput while maintaining comparable accuracy for all {\it slim, medium} and {\it deep} CNNs without using any re-training techniques.

\section{Low Precision Floating-point Quantization}
\label{section:LFP}

In this section, we present the details of our proposed low precision floating-point (LPFP) quantization method, including the quantization process, data flow in processor and quantization results.

\begin{figure}[t]
\centering
\includegraphics[width=3.5in]{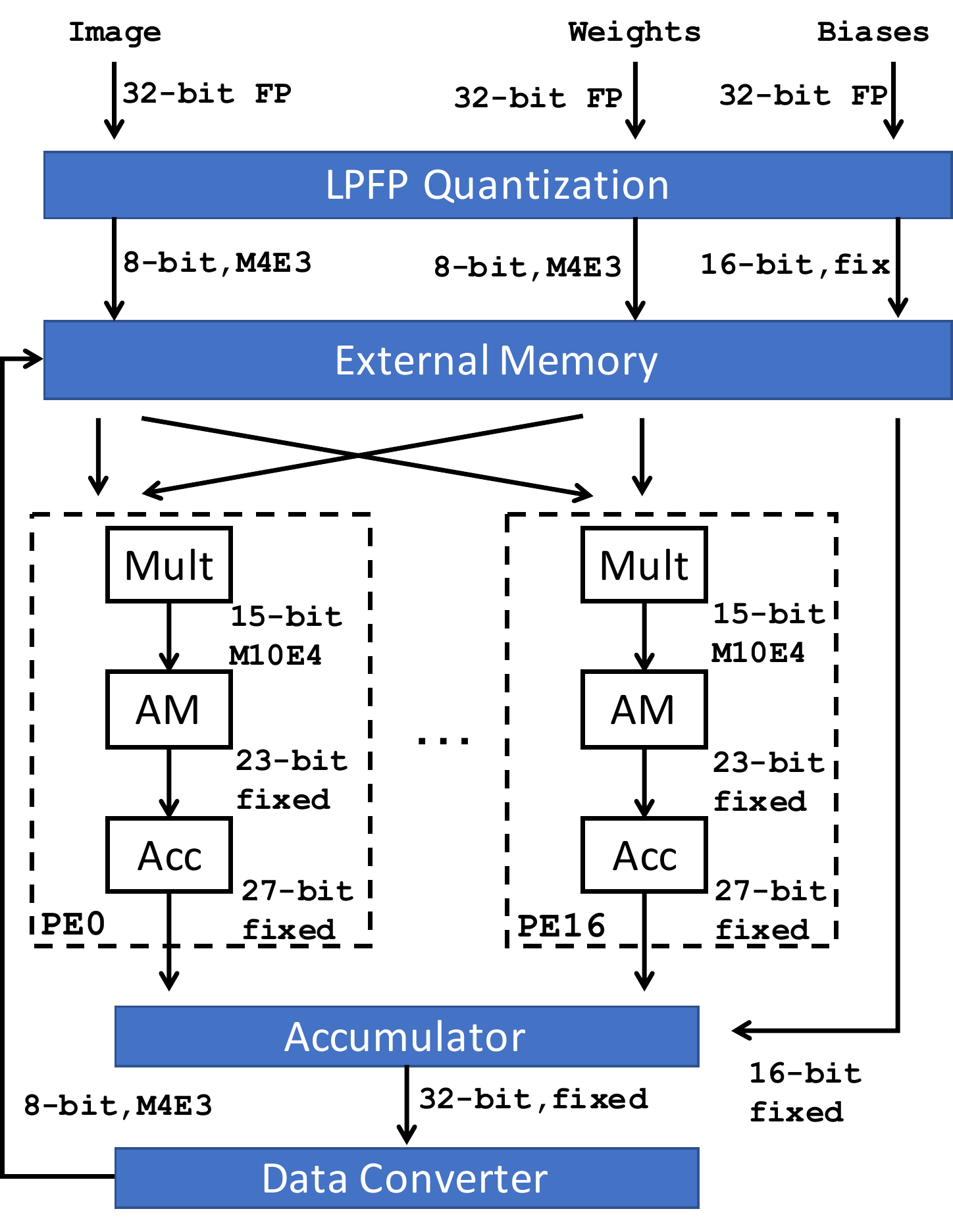}
\caption{The data flow in our processor with $M4E3$ data format as an example (FP: floating-point, Mult: LPFP multiplier, AM: alignment module, Acc: accumulator, DC: data converter).}
\label{fig:data_flow}
\end{figure}

\subsection{Quantization Process}
\label{subsection:quantization_process}

Our proposed LPFP quantization method is applied to both activations and weights. The quantization function is defined as follows: 
\begin{equation}
\label{equation:quantization}
    V_{lfp} = quan(V_{fp32} \times 2^{sf}, MIN_{lfp}, MAX_{lfp}),
\end{equation}
where $V_{lfp}$ and $V_{fp32}$ denote the decimal values represented by LPFP and traditional single floating-point format, respectively; $MIN_{lfp}$ and $MAX_{lfp}$ indicate the minimal and maximal numbers represented by LPFP, and $sf$ is the scaling factor which is used to better fit the data into the dynamic range of LPFP. The $quan$ function in Eq. ~(\ref{equation:quantization}) rounds the data to the nearest value with saturation considered, formulated as
\begin{equation}
\label{equation:round}
    quan(x, MIN, MAX) = 
    \begin{cases}
        MIN &x <= MIN \\
        MAX &x >= MAX \\
        round(x) &\text{otherwise}
    \end{cases}
    ,
\end{equation}
where $MIN$ and $MAX$ are the minimal and maximal values, respectively. 

The mean square error (MSE) of the values before and after quantization is used as the metric to evaluate the quantization error, illustrated as:
\begin{equation}
\label{equation:mse}
    MSE = \frac{1}{N}\sum_{i=0}^N (V_{lfp} / 2^{sf} - V_{fp32})^2,
\end{equation}
where $N$ denotes the amount of data.

As illustrated from Eq. ~(\ref{equation:quantization}) to (\ref{equation:mse}), MSE is influenced by the data format of LPFP and the scaling factor ($sf$). We will find an optimal combination of LPFP data format and scaling factor for least MSE. In this paper, we assume {\bf the same data format for a CNN and a same scaling factor for each layer}. This assumption can be removed as needed. Furthermore, we choose to use a same optimized data format for all test cases in our experiments, while the problem formulation is to decide a data format for each CNN. 

\subsection{Data Flow in Processor}
\label{subsection: dataflow}

The data flow to run inference of a quantized network in our processor is shown in Figure~\ref{fig:data_flow}. In order to explicitly illustrate the data flow, we list the bit width in each step with $M4E3$ data format as an example. All the input image, weights and biases are represented by 32-bit floating-point. In our processor, the raw input image which indicates the input of the first layer and all the weights are quantized with $M4E3$ data format and stored in external memory, while biases are quantized to 16-bit fixed-point to reduce quantization error. Multiplications are performed with the quantized image and weights, and the 15-bit floating-point ($M10E4$) products are converted to 23-bit fixed-point without any precision loss. In this way, all the accumulation can be done in fixed-point accumulators, which consumes fewer resources in FPGA than floating-point accumulators. The final outputs in each output channel are converted to $M4E3$ floating-point again (and stored in the external memory) before being used by another CNN layer. In the data flow, only the final data conversion step introduces bit truncation and precision loss. However, the precision loss introduced by the final step has little impact on the final accuracy and is validated in Subsection~\ref{subsection:quantization_results} with comprehensive experimental results. 

\subsection{Quantization Results}
\label{subsection:quantization_results}

\subsubsection{Experiment Setup}
\label{subsubsection:expr_environment}

We implement our LPFP quantization method with C language based on the Darknet framework \cite{darknet}, and the inference process of the quantized network follows the same data flow as that in our processor illustrated in Figure~\ref{fig:data_flow}. The validation accuracy with single center-crop is then evaluated via the ImageNet validation set (50,000 labelled images) \cite{imagenet}. Our quantization process is run on an Intel (R) Core (TM) i9-7960X CPU working under 2.86GHz, while the evaluation process is run on a Nvidia TITAN Xp GPU. Six representative CNNs including the {\it slim, medium} and {\it deep} CNNs are evaluated, as listed in Table~\ref{table:benchmark}. 

\begin{figure}[t]
\centering
\includegraphics[width=3.5in]{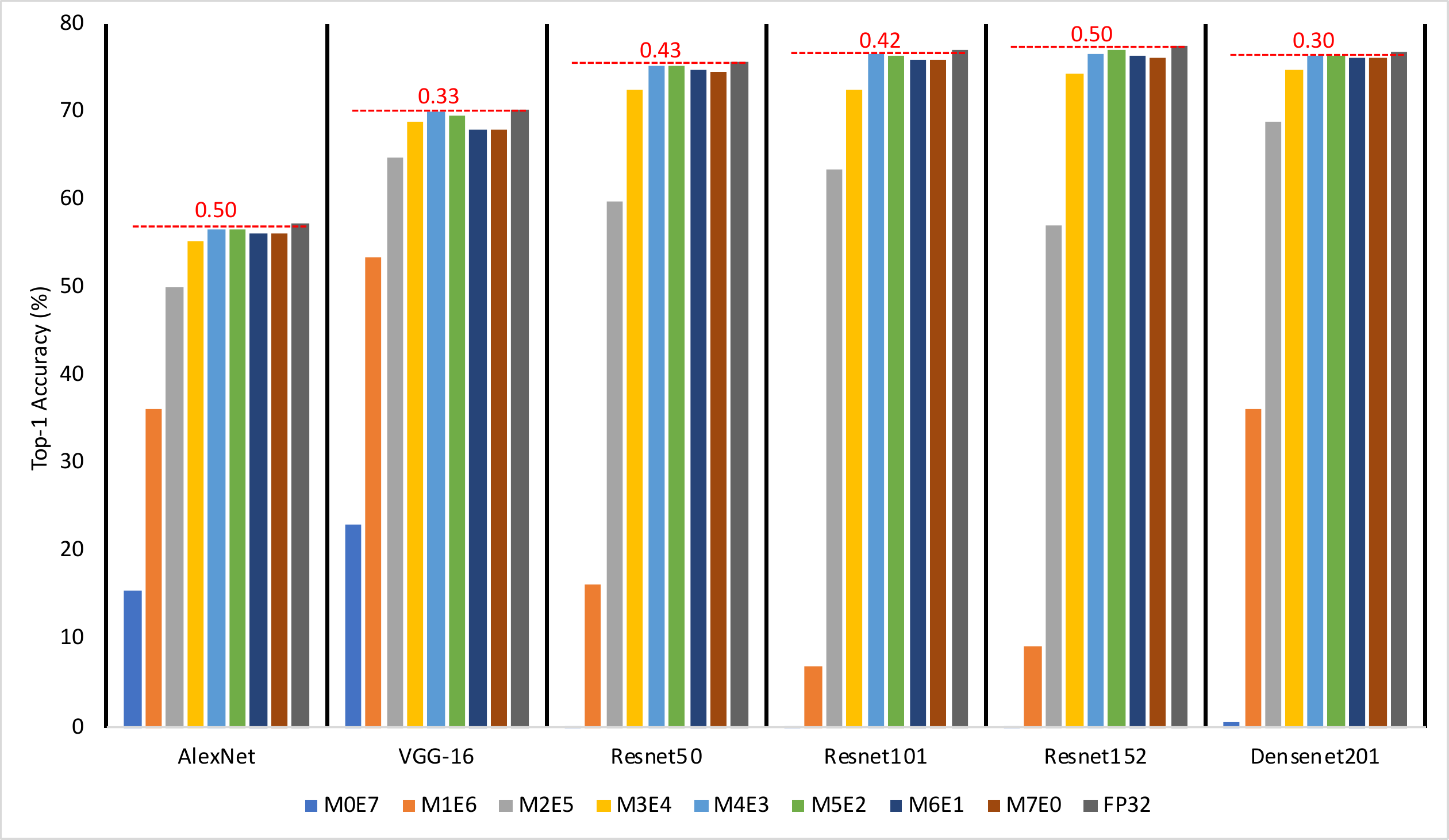}
\caption{Top-1 accuracy for different (mantissa, exponent) combinations with respect to different CNNs.}
\label{fig:top_1}
\end{figure}

\subsubsection{8-bit Quantization}
\label{subsubsection:8_quan_res}

\begin{figure}[t]
\centering
\includegraphics[width=3.5in]{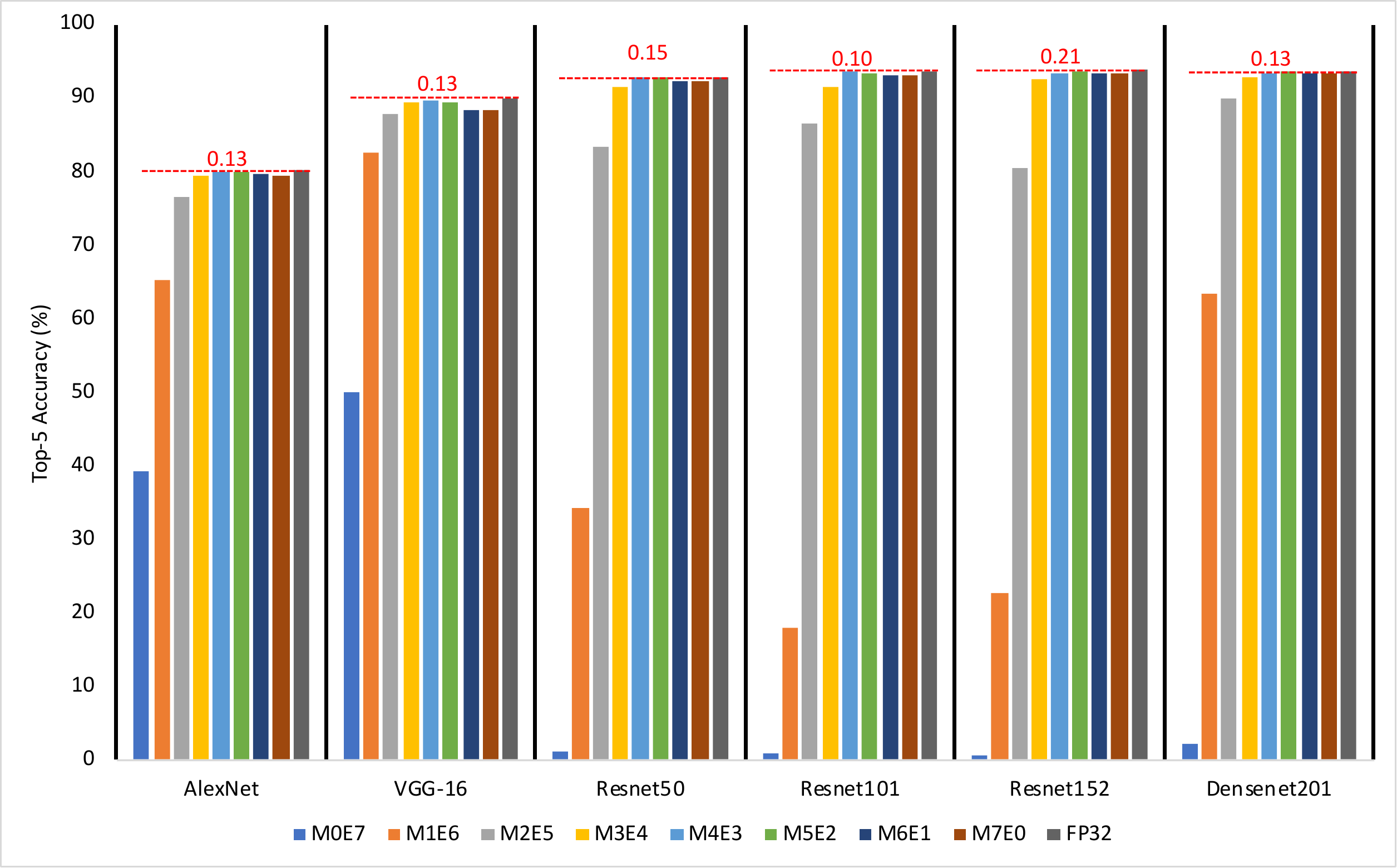}
\caption{Top-5 accuracy for different (mantissa, exponent) combinations with respect to different CNNs.}
\label{fig:top_5}
\end{figure}

The detailed validation accuracies on the quantized networks with 8-bit floating-point data format are shown in Figures~\ref{fig:top_1} and ~\ref{fig:top_5}. We emulate all 8 different (mantissa, exponent) combinations to validate the top-1 and top-5 accuracy of the quantized CNNs, and the 32-bit floating-point results are included as the baseline.

In Figures~\ref{fig:top_1} and ~\ref{fig:top_5}, the dashed lines illustrate the 32-bit floating-point baseline, while the values above the dashed lines are the accuracy loss compared with the baseline. We can see that our LPFP quantization approach can maintain comparable top-1 and top-5 accuracy to the baseline. On average, the top-1 and top-5 accuracy loss is within 0.5\% and 0.3\% compared with the full precision results, respectively. Particularly, $M5E2$ always achieves the highest accuracy compared with the other cases. Data formats with more than or equal to 3-bit mantissa all have a low accuracy loss for all the six CNNs, while those with less than 3-bit mantissa can hardly find accurate results. We also compare our proposed approach with the fixed-point situation, marked as $M7E0$ in the figures ($M7E0$ means 1-bit sign, 7-bit mantissa and no exponent, exactly fixed-point). As shown in Figures~\ref{fig:top_1} and ~\ref{fig:top_5}, $M4E3$ and $M5E2$ outperform the fixed-point for all six benchmarks. 

\subsubsection{Lower Bit Width Quantization}
\label{subsubsection:lower_quan_res}

\begin{figure}[t]
\centering
\includegraphics[width=3.5in]{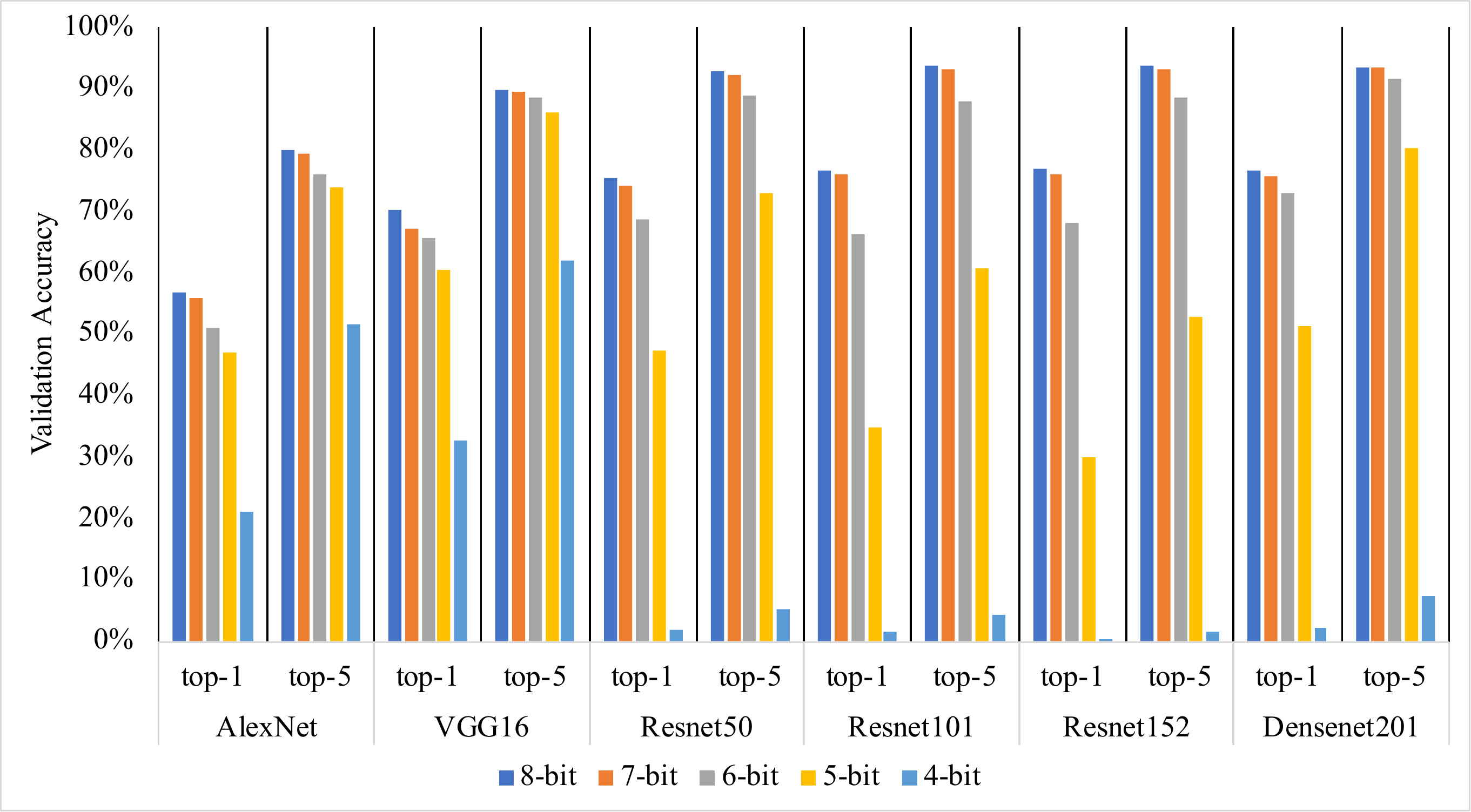}
\caption{Top-1 and top-5 accuracies for different bit width with respect to different CNNs.}
\label{fig:lfp8_4}
\end{figure}

We further reduce the bit width from 8-bit to 4-bit and also evaluate the top-1 and top-5 accuracy of the quantized networks. We pick the best (mantissa, exponent) combination for each data format and the results are shown in Figure~\ref{fig:lfp8_4}. We can see that both the top-1 and top-5 accuracy decrease when lower bit length is utilized to represent the weights and activations of CNNs. Particularly, the average top-5 accuracy degradations for 7-bit and 6-bit are 0.8\% and 4.2\%, respectively. However, the accuracy drops dramatically when the bit width decreases to less than 6 bits, which means our LPFP quantization approach can hardly find accurate results without any re-training process.

\subsubsection{Comparison with the Prior Quantization Strategies}
\label{subsubsection:comp_soa}

$M4E3$ and $M5E2$, which achieve the two best accuracies among all the test cases, are also compared with five typical approaches. We report both the top-1 and top-5 accuracy for all six benchmarks in Table~\ref{table:acc_cmp}, where "-" indicates no reported results in the literatures. We use the normalized top-1 accuracy in Table~\ref{table:acc_cmp} for the approaches proposed by ARM \cite{ARM_mixed} and Xilinx \cite{xilinx_float} as reported in their paper. The top-1 and top-5 accuracy in Table~\ref{table:acc_cmp} show that our LPFP quantization method without any re-training can outperform the literatures in most cases. Moreover, besides the approach proposed by Nvidia \cite{nvidia_fix}, our method is the only one that can reach \textit{deep} networks.

\begin{table*}
\centering
\caption{Accuracy comparison between $M4E3$, $M5E2$, references and FP32. "-" means no reported results. Results for ARM and Xilinx are normalized top-1 accuracy as reported in their papers without circuit implementations and we convert them into actual accuracy, while others are based on circuit implementations.}
  \begin{tabular}{c|cc|cc|cc|cc|cc|cc}  \hline
    \multirow{2}{*}{}             &\multicolumn{12}{c}{Top-1 Accuracy (\%) \quad Top-5 Accuracy (\%) for each network} \\ \cline{2-13}
                                  &\multicolumn{2}{c|}{AlexNet}   &\multicolumn{2}{c|}{VGG16}     &\multicolumn{2}{c|}{ResNet50} 
                                  &\multicolumn{2}{c|}{ResNet101} &\multicolumn{2}{c|}{ResNet152} &\multicolumn{2}{c}{DenseNet201}     \\ \hline
    Angle-Eye \cite{angle_eye}    &-          &-                  &67.72      &88.06              &-          &-
                                  &-          &-                  &-          &-                  &-          &-                       \\ \hline
    Nvidia \cite{nvidia_fix}      &57.05      &80.06              &70.84      &-                  &73.10      &91.06
                                  &74.40      &91.73              &74.70      &91.78              &-          &-                       \\ \hline
    ARM \cite{ARM_mixed}          &56.71      &-                  &70.38      &-                  &-          &-
                                  &-          &-                  &-          &-                  &-          &-                       \\ \hline
    Xilinx \cite{xilinx_float}    &-          &-                  &-          &-                  &75.80      &-
                                  &-          &-                  &-          &-                  &-          &-                       \\ \hline
    BFP \cite{bfp_fpga}           &-          &-                  &68.32      &-                  &72.76      &-
                                  &-          &-                  &-          &-                  &-          &-                       \\ \hline
    FP32 (Baseline)               &57.28      &80.18              &70.38      &89.81              &75.80      &92.90
                                  &77.10      &93.70              &77.60      &93.83              &76.85      &93.62                   \\ \hline
    {\bf Ours ($M4E3$)}           &{\bf 56.69}&{\bf 79.99}        &{\bf 70.05}&{\bf 89.68}        &{\bf 75.25}&{\bf 92.75}
                                  &{\bf 76.68}&{\bf 93.60}        &{\bf 76.79}&{\bf 93.44}        &{\bf 76.40}&{\bf 93.43}             \\ \hline
    {\bf Ours ($M5E2$)}           &{\bf 56.77}&{\bf 80.05}        &{\bf 69.74}&{\bf 89.49}        &{\bf 75.37}&{\bf 92.71}
                                  &{\bf 76.43}&{\bf 93.33}        &{\bf 77.05}&{\bf 93.62}        &{\bf 76.55}&{\bf 93.49}             \\ \hline
  \end{tabular}
  \label{table:acc_cmp}
\end{table*}

\section{Processor Architecture}
\label{section:proc_arch}

In this section, we discuss in detail the architecture of the processor, which efficiently supports the inference process of quantized networks for various CNNs.

\subsection{Overview}
\label{subsection:overview}

\begin{figure}[t]
\centering
\includegraphics[width=3.5in]{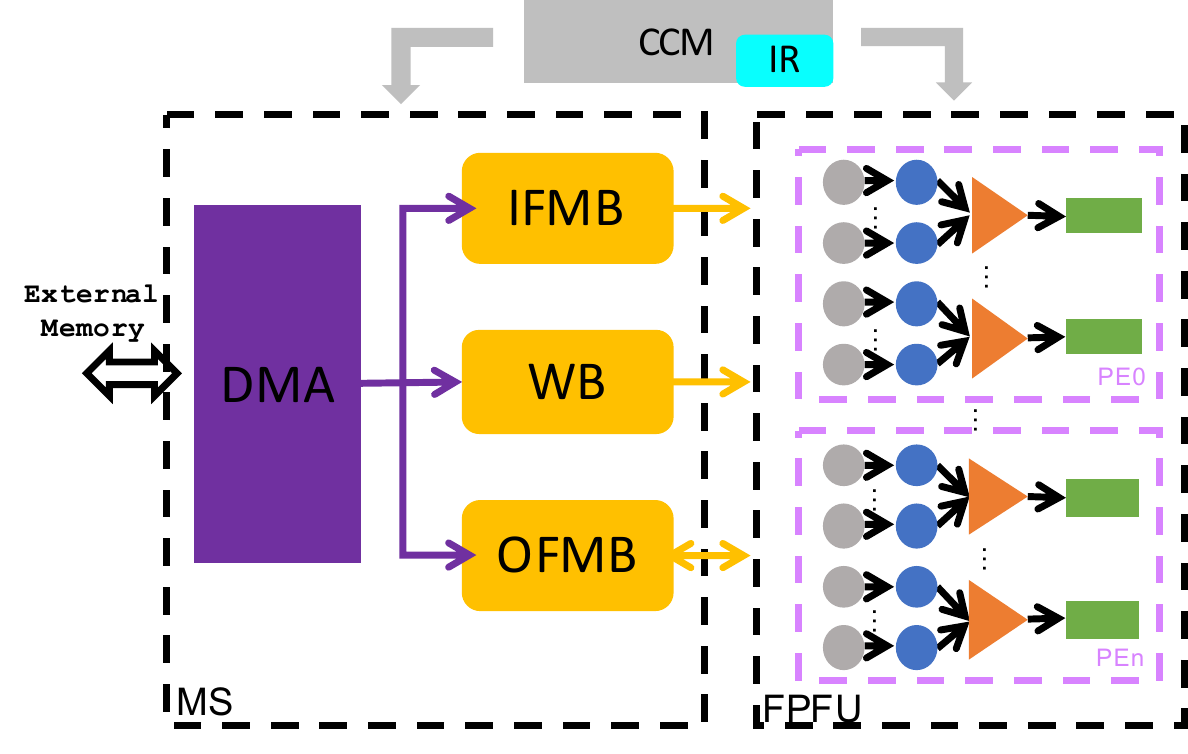}
\caption{The overall architecture of proposed processor.}
\label{fig:overview}
\end{figure}

The overall architecture of the proposed processor is depicted in Figure~\ref{fig:overview}. A floating-point function unit (FPFU), which is composed of multiple processing elements (PEs), is developed to compute the outputs of a layer in parallel. The PE, which is the key component of FPFU, is designed to efficiently perform dot product with LPFP data format. The on-chip memory system (MS) consists of three buffers, {\it e.g.,} input feature map buffer (IFMB), weight buffer (WB) and output feature map buffer (OFMB). All these three buffers are ping-pong architecture to hide the communication time between on-chip and off-chip memories through direct memory access (DMA) module. The central control module (CCM) is designed to arbitrate between different modules. Moreover, the CCM decodes various instructions stored in the instruction RAM (IR) into detailed signals for other modules. 

\subsection{Floating-point Function Unit}
\label{subsection:FPFU}

FPFU, which is constructed by multiple PEs, is designed to perform convolution in LPFP data format efficiently for performance gain and power reduction. Different parallel computation patterns, including parallel in input feature maps, parallel in output feature maps and parallel in both input and output feature maps, are developed in FPFU and are discussed in the following paragraphs. FPFU receives activations and weights from IFMB and WB, respectively, and distributes the activations and weights to different PEs to perform convolution according to the control signals decoded by CCM.

\subsubsection{Architecture of PE}
\label{subsubsection:arch_pe}

\begin{figure}[t]
\centering
\includegraphics[width=3.5in]{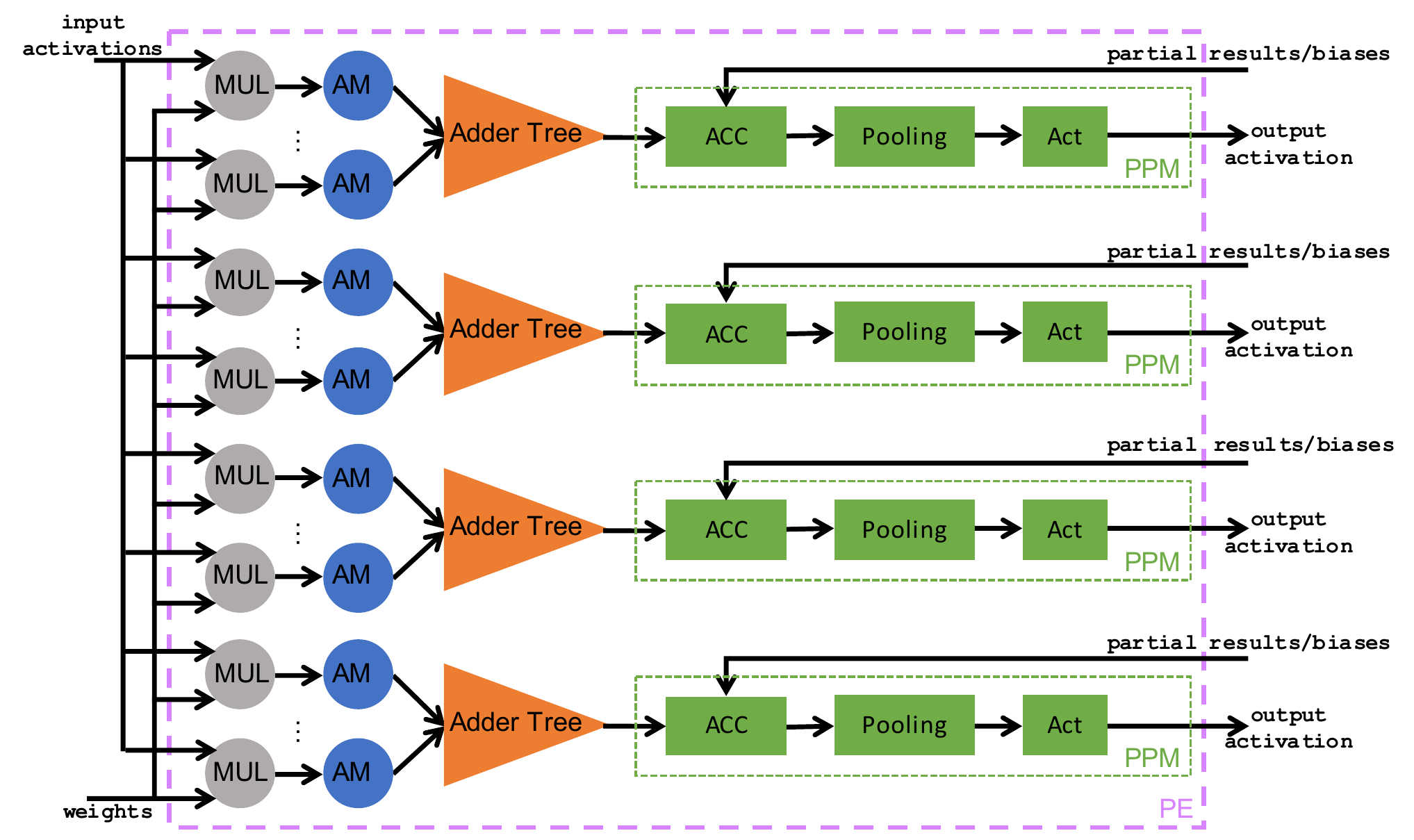}
\caption{The architecture of a PE. MUL: LPFP multiplier, AM: alignment module, ACC: accumulator, Act: activation.}
\label{fig:PE}
\end{figure}

The PE is designed as a fully pipelined data-flow-based architecture, as shown in Figure~\ref{fig:PE}. Once a PE receives two vectors, it distributes the data to $N_m$ multipliers inside the PE, whose full precision floating-point results are transferred into the alignment module (AM). The full precision floating-point products are aligned and converted to fixed-point numbers without any bit truncation. The aligned products are then fed into four fixed-point adder trees to finalize four dot product processes in parallel, which indicates the feed-forward process of four pixels in two output channels (see details in Subsubsection~\ref{subsubsection:para_calc}). The accumulation of partial results (including bias), pooling and activation processes are performed in series inside the post process module (PPM). 

The multipliers in each PE are developed for LPFP, which are represented with scientific notation in the sign-and-magnitude format, as illustrated in Eqs. ~(\ref{equation:normal}) and ~(\ref{equation:subnormal}). The multiplication of two LPFP numbers is then divided into three fixed-point components: (1) XOR of the signs; (2) multiplication of mantissas; (3) addition of exponents. Take the $MaEb$ format as an example. An $a$-bit unsigned MAC and a $b$-bit unsigned adder are needed. Although the multiplication of mantissas should be $a+1$-bit considering the first hidden bit of mantissas -- "1" for normal numbers and "0" for subnormal numbers -- we design the $a$-bit MAC to perform the $a+1$-bit multiplication to improve per DSP throughput (see details in Subsection~\ref{subsection:impl_details}). Meanwhile, the exponent bias $E_b$ is not included during addition, because the $E_b$ is the same for all the numbers in one CNN as we assume, and we can address this at the last step to simplify the adders.

\subsubsection{Parallel Computation Pattern}
\label{subsubsection:para_calc}

\begin{figure}[t]
\centering
\includegraphics[width=3.5in]{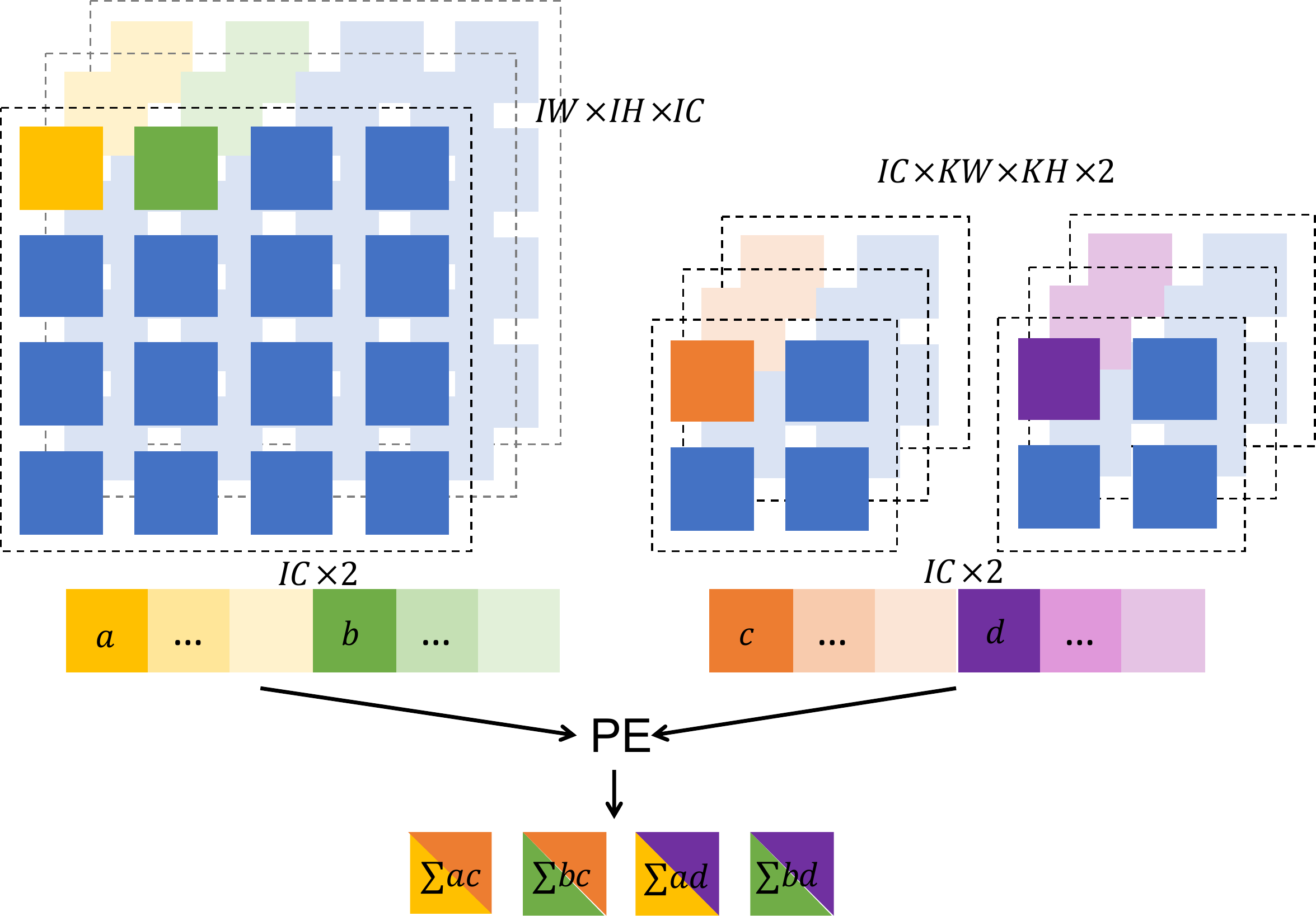}
\caption{Parallel computation pattern in one PE.}
\label{fig:calc_pattern}
\end{figure}

During the convolution process, each pixel in one output channel is calculated as

\begin{equation}
\label{equation:conv}
y_i = \sum_{k=0}^{KW\times KH}\sum_{ic=0}^{IC}x_{k,ic}w_{k,ic} + b_i,
\end{equation}
where $IC$ indicates the number of input channel, $KW$ and $KH$ denotes the width and height of the kernel, and $x, y, w$ and $b$ are input activation, output activation, weight and bias, respectively. In our implementation on FPGA, we implement 4 LPFP multipliers with one DSP slice, which follows the pattern: $(a+b)\times(c+d)=ac+bc+ad+bd$ (see details in Subsection~\ref{subsection:impl_details}). Therefore, each PE is designed to process convolution in two output channels in parallel, and in each output channel, it will calculate the convolutional results of two pixels at the same time, as shown in Figure~\ref{fig:calc_pattern}. To be specific, in the first cycle, the first pixel in $IC$ input channels and the first value in the corresponding kernels are fed into the PE, marked with $a$ and $c$ in Figure~\ref{fig:calc_pattern}, respectively. To follow the computation pattern in these four multipliers, the second pixel in $IC$ input channels (marked with $b$), and the corresponding kernels to calculate the pixel in another output channel (marked with $d$) are also fed into the PE. In this way, $a$ and $b$ are reused to produce the pixels in different output channels, while $c$ and $d$ are reused to produce the pixels in different positions of the same output channel. After $KW \times KH$ cycles, four convolution results are produced by one PE. 

As illustrated in Subsubsection~\ref{subsubsection:arch_pe}, $N_m$ multipliers are used in each PE, and $IC$ is designed to be $N_m / 4$. In this way, $N_m / 4$ input channels are calculated in parallel in each PE. With the corresponding weights and biases, 2 pixels in 2 output channels are calculated in parallel. When the number of input channels is larger than $N_m / 4$ and/or when the number of pixels in each output channel is larger than 2 and/or when the number of output channels is larger than 2, multiple rounds of computation are needed in series to finalize the convolution. In order to further increase the parallelism, we use $N_p$ PEs in the FPFU. In different PEs, we can feed in different pixels in input feature maps and weights to perform different parallel computation pattern. For example, the $N_p$ PEs can share the same input feature map and use different weights to parallelize the computation in output channels, or the $N_p$ PEs can share the same weights and use different input feature maps to parallelize the computation in input channels. The $N_m$, $N_p$ and the parallel computation pattern are decided by considering the CNNs, the throughput and the bandwidth requirement. This will be explained with experiments in Subsubsection~\ref{subsubsection:para_exp}.

\subsection{Memory System}
\label{subsection:on_chip_mem}

Following the computation pattern in PE, the IFMB and the WB are set to provide $N_m / 2$ LPFP input activations and weights to each PE at every cycle, respectively, while the OFMB needs to save 4 output activations from each PE at every cycle. Although each pixel in the output feature map is represented with LPFP data format, we keep the intermediate results with 16-bit precision to reduce accuracy loss. In this way, the bit width of OFMB for each PE is set to 64 bits. As the input activations and/or weights can be shared by different PEs according to different computation patterns, we define $P_{ifm}$ and $P_{ofm}$ ($P_{ifm} \times P_{ofm} = N_p$) to indicate the parallelisms in input feature map and output feature map, respectively. In this definition, $P_{ifm}$ indicates that we have $P_{ifm}$ PE groups where the same weights are shared during calculation, while in each PE group, $P_{ofm}$ PEs share the same input activations. Therefore, the bit width for IFMB, WB and OFMB are $N_m / 2 \times P_{ifm}\times BW$, $N_m / 2 \times P_{ofm}\times BW$ and $64N_p$, respectively, where $BW$ denotes the bit width of LPFP data format. 

The parameters $N_m$, $P_{ifm}$ and $P_{ofm}$ are decided to trade off between the throughput, bandwidth requirement and resource utilization. The sizes of the three buffers also determine the throughput and resource utilization. Previous proposed work applied large enough buffers to store all the activations or weights for one layer \cite{EIE} to avoid costly off-chip memory access. However, such designs incurred large area and unscalability for larger and deeper CNNs. In our processor, we trade off among the throughput, bandwidth requirement, resource utilization and scalability, and employ the smallest sizes which can hide the DMA communication time. In our implementation on FPGA, we use block RAM to deploy IFMB and OFMB, while we use distributed RAM to deploy WB, as distributed RAM can provide higher bandwidth than block RAM. During inference on our processor, only when all the input feature maps have been processed and reused, or all the weights have been processed and reused, or OFMB is full, will the off-chip memory be accessed for loading new input feature maps, loading new weights or storing output feature maps, respectively.

\subsection{Central Control}
\label{subsection:central_ctrl}

The CCM is designed to arbitrate among different modules and control the whole execution process. First, CCM decodes the instructions from IR efficiently and sets the corresponding control registers. Second, different modules are activated according to the control registers and the status of each module is monitored by the control registers as well. Finally, the CCM decides when to fetch the next instruction from the feedback of the control registers. We also design a compiler to generate the block-level instructions.

\begin{figure}[t]
\centering
\includegraphics[width=3.5in]{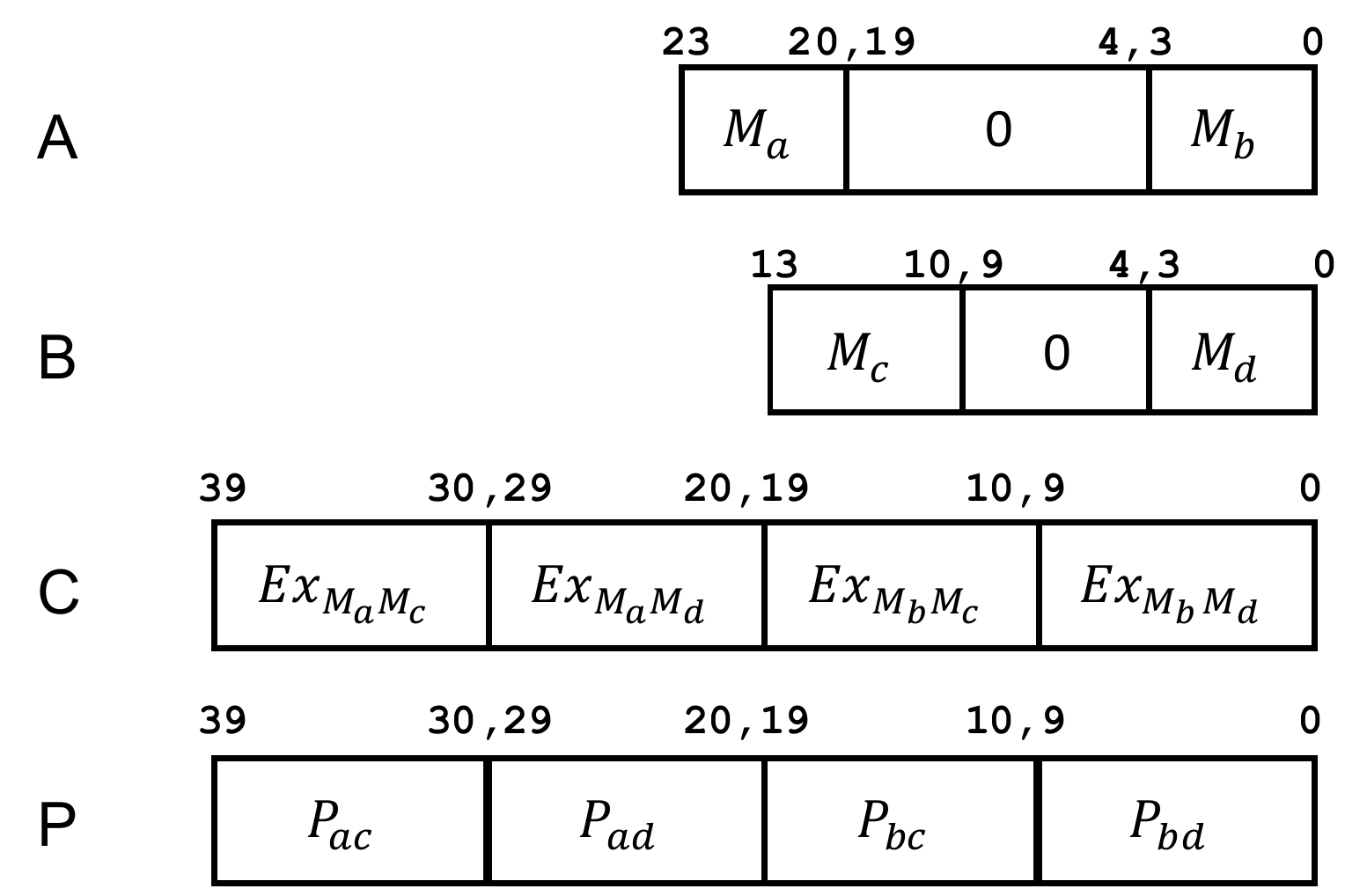}
\caption{Data format of the DSP to implement four 4-bit MACs. $M_{a}, M_{b}, M_{c}$ and $M_{d}$: the mantissas of LPFP data $a, b, c$ and $d$, respectively; $Ex_{M_{a}M_{c}}, Ex_{M_{a}M_{d}}, Ex_{M_{b}M_{c}}$ and $Ex_{M_{b}M_{d}}$: the extra term expressed as $Ex_{M_{a}M_{c}}=1.M_{a}+0.M_{c}$; $P_{ac}, P_{ad}, P_{bc}$ and $P_{bd}$: the mantissas of the product of two LPFP data expressed as $P_{ac}=1.M_{a} \times 1.M_{c}$.}
\label{fig:4mul}
\end{figure}

\section{Evaluation}
\label{section:eva}

In this section, the environment setup of our evaluation is first introduced, then the implementation details and comprehensive experimental results are provided.

\subsection{Environment Setup}
\label{subsection:exp_meth}

Our processor is implemented on the KC705 evaluation board, which includes a Xilinx Kintex-7 XC7K325T FPGA and a 1GB DDR3 module. First, we explore the parallel computation patterns to find the optimal parameters to best fit the FPGA on KC705. Second, with these parameters, the processor is described in Verilog-HDL, and synthesized and implemented with the Xilinx Vivado 2018.2 Design Suite. Finally, we evaluate the throughput and per DSP throughput of running different networks (shown in Table~\ref{table:benchmark}) on our processor, and the results are compared with two prior accelerators \cite{16_fix_2, rna}. The Intel (R) Core (TM) i9-7960X CPU under 2.86GHz working frequency and the Nvidia TITAN Xp GPU with a 12GB DDR5 are also used for comparison. More comprehensive experimental results on VGG16 and YOLO are compared with latest FPGA accelerators \cite{16_fp, 16_fix_1, 16_fix_2, angle_eye, rna, bfp_fpga, fix_tinyyolov2}. 

\subsection{Implementation Details}
\label{subsection:impl_details}

We use the $M4E3$ data format for FPGA implementation in this paper for two reasons. First, $M4E3$ achieves the top two best validation accuracies among all the LPFP (mantissa, exponent) combinations we tested (see Subsection~\ref{subsection:quantization_results}). Particularly, the average top-1 and top-5 accuracy loss of $M4E3$ compared with 32-bit floating-point are 0.53\% and 0.19\%, respectively. Second, $M4E3$ only needs a 4-bit fixed-point MAC and a 3-bit fixed-point adder, resulting in fewer resources on FPGA than $M5E2$. To be specific, four 4-bit fixed-point MACs can be implemented inside one DSP48E1 slice in XC7K325T FPGA.

In order to clearly explain the way to implement four MACs with one DSP48E1 slice, we take the multiplication of two normal numbers ($X$ and $Y$) as an example. The mantissa of the product can be explained as:

\begin{equation}
\label{equation:lfp_mul}
    \begin{split}
        Prod &=1.M_x \times 1.M_y \times 2^{(2-E_x-E_y)} \\
             &=(0.M_x \times 0.M_y + (1.M_x + 0.M_y)) \times 2^{(2-(E_x+E_y))},
    \end{split}
\end{equation}
where $M_x, M_y, E_x$ and $E_y$ are the mantissas and exponents of $X$ and $Y$, respectively. In Eq. ~(\ref{equation:lfp_mul}), the term $0.M_x \times 0.M_y + (1.M_x + 0.M_y)$ is performed with a 4-bit unsigned fixed-point MAC and the term $E_x+E_y$ is performed with an extra 3-bit unsigned fixed-point adder. As the DSP48E1 slice can be implemented as a MAC followed by $P=A\times B+C$ (where the maximal bit width of $A, B$ and $C$ are 25, 18 and 48, respectively), we add blank bits to the three inputs to fully utilize the functionality of DSP48E1, as shown in Figure~\ref{fig:4mul}. During the calculation process, the dot position is kept at the right most position. That is, the terms $0.M_x$ and $0.M_y$ are converted to 4-bit integers, while the extra term $1.M_x + 0.M_y$ is converted to 10-bit integers to make sure that no overlap occurs. In this way, with a few LUTs and FFs to perform additions of the exponents and the extra term $1.M_x + 0.M_y$, four multiplications with $M4E3$ data format can be carried out in on DSP slice (see Table~\ref{table:mul_dsp}), thus dramatically increasing the per DSP throughput.

\subsubsection{Parallel Exploration}
\label{subsubsection:para_exp}

\begin{figure}[t]
\centering
\includegraphics[width=3.5in]{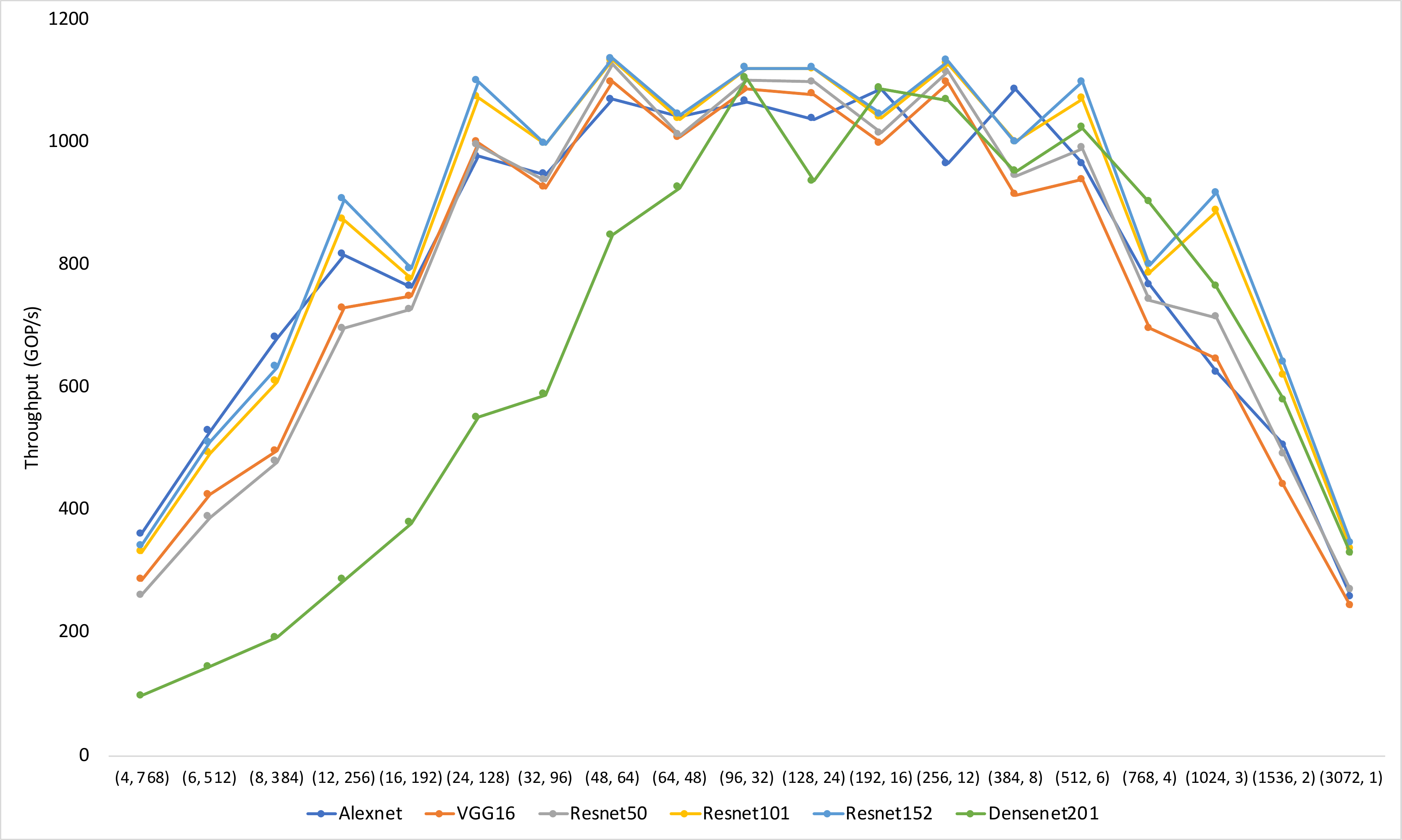}
\caption{Throughput for different CNNs with respect to different $N_m$ and $N_p$ combinations.}
\label{fig:nmnp}
\end{figure}

\begin{figure}[t]
\centering
\includegraphics[width=3.5in]{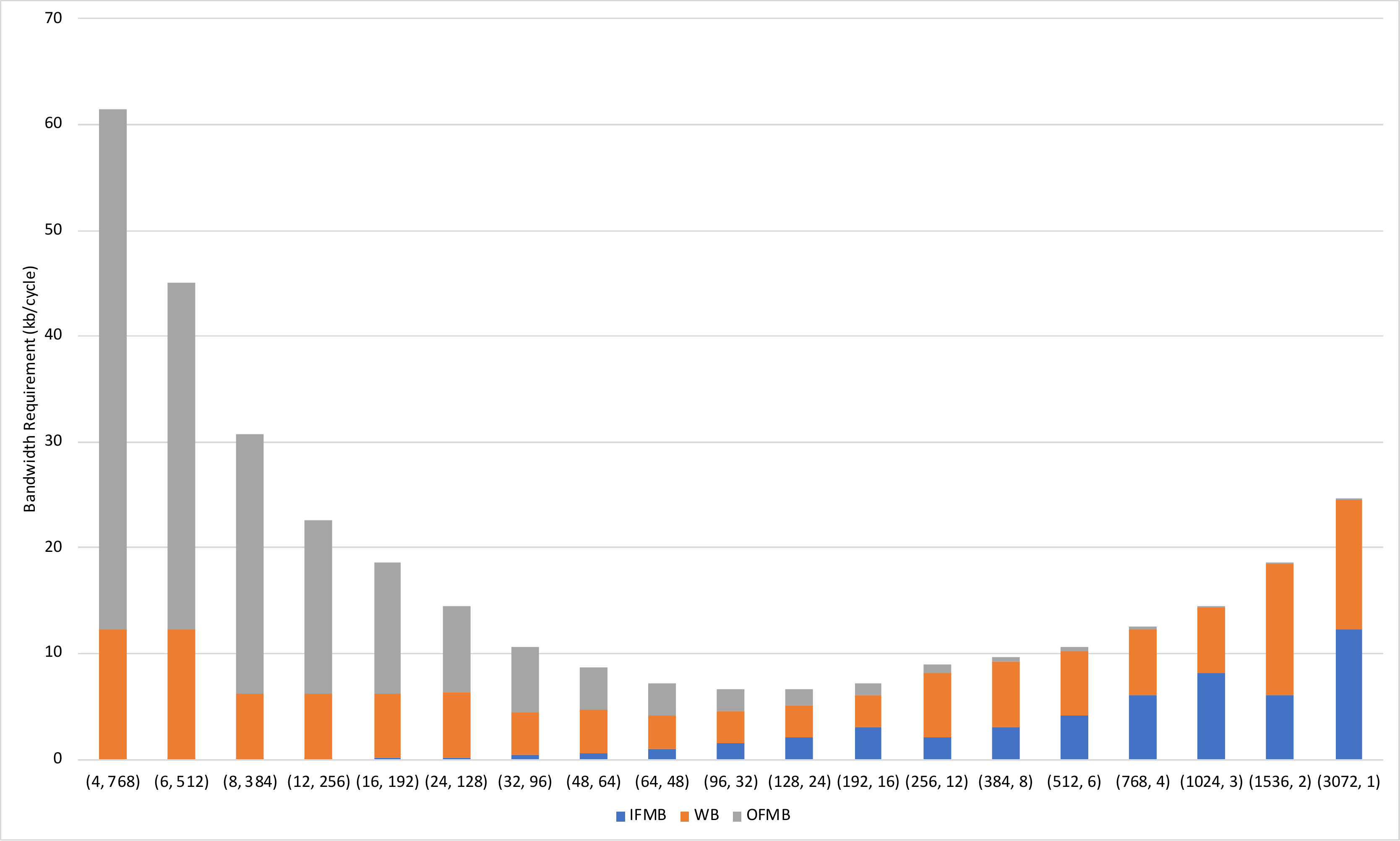}
\caption{Bandwidth requirement with respect to different $N_m$ and $N_p$ combinations.}
\label{fig:bw_req}
\end{figure}

Since one DSP slice is divided into four 4-bit LPFP MACs in our implementation, the parameters should meet the requirement that $N_m \times N_p=4 \times \# of DSP$. Considering the resources of XC7K325T FPGA, we set the targeted number of DSP as 768, which accounts for 91.43\% of the available DSPs. We then evaluate the throughput for different CNNs and the bandwidth requirement with respect to different $N_m$ and $N_p$ combinations as shown in Figures~\ref{fig:nmnp} and ~\ref{fig:bw_req}, respectively. We also explore different combinations of the parameters $P_{ifm}$ and $P_{ofm}$, and only depict the $P_{ifm}$ and $P_{ofm}$ for achieving the optimal throughput and minimal bandwidth requirement in Figures~\ref{fig:nmnp} and~\ref{fig:bw_req}.

In general, when $N_m$ keeps increasing, the throughput first increases and then decreases when it reaches the peak. The small $N_m$ and large $N_p$ indicate that more output channels are calculated in parallel while large $N_m$ and small $N_p$ mean more input channels are calculated in parallel. When $N_m$ is larger than the total number of input channel (denoted as $IC$), only $IC$ multipliers are used while the rest are wasted, resulting in a low throughput. This is the same for large $N_p$, and the peak throughput comes from balanced $N_m$ and $N_p$. For different CNNs, the peak throughput comes from different $N_m$ and $N_p$ combinations due to different network configurations. For example, DenseNet201 has lots of inception layers, which concatenate layers with small output channels ({\it e.g.,} 32) to form layers with large input channels ({\it e.g.,} 1568). In this case, larger $N_m$ and smaller $N_p$ incur fewer wasted computations and lead to higher throughput. From Figure~\ref{fig:nmnp}, we can see that the combination of $N_m=96$ and $N_p=32$ results in an optimal throughput for all cases on average.

The bandwidth requirement is extremely high when $N_p$ is large. This is because larger $N_p$ indicates more parallel computations in output channels. Moreover, OFMB is designed to store 16-bit intermediate results, which also lead to higher bandwidth requirement with larger $N_p$. The total bandwidth requirement decreases when $N_p$ decreases, and then increases again since larger $N_m$ needs more bandwidth to load input activations and weights. The smallest bandwidth requirement comes when we have a balanced combination of $N_m$ and $N_p$. As concluded from Figure~\ref{fig:bw_req}, the optimal combinations are $N_m=96, N_p=32$ and $N_m=128, N_p=24$. Take the case for optimal throughput, we set $N_m=96$ and $N_p=32$ in this implementation.

\subsection{Experimental Results}
\label{subsection:exp_res}

\subsubsection{Resource Utilization}

Given the parameters that $N_m=96$ and $N_p=32$, the detailed post-implementation resource utilization under 200MHz working frequency is listed in Table~\ref{table:resource}.

\begin{table}
  \centering
  \caption{Resource utilization in XC7K325T.}
  \begin{tabular}{c|c|c|c|c|c}  \hline
    Resource                &LUT             &LUTRAM        &FF                &BRAM         &DSP         \\ \hline
    Used                    &154625          &7860          &180561            &234.5        &768         \\ \hline
    Available               &203800          &64000         &407600            &445          &840         \\ \hline
    Utilization             &75.87\%         &12.28         &44.30\%           &52.70\%      &91.43\%     \\ \hline
  \end{tabular}
  \label{table:resource}
\end{table}

\begin{table*}
  \centering
  \caption{Comparison between Intel i9 CPU, Nvidia TITAN Xp GPU, existing accelerators and our processor with respect to different CNNs. "-" means no reported results.}
  \begin{tabular}{c|cc|cc|cc|cc|cc|cc}  \hline
       \multirow{2}{*}{}&\multicolumn{12}{c}{Throughput (GOPS) \quad per DSP throughput (also called DSP efficiency, unit: GOPS/DSP) for each network} \\\cline{2-13} 
                                    &\multicolumn{2}{c|}{AlexNet}      &\multicolumn{2}{c|}{VGG16}          &\multicolumn{2}{c|}{ResNet50}  
                                    &\multicolumn{2}{c|}{ResNet101}    &\multicolumn{2}{c|}{ResNet152}      &\multicolumn{2}{c}{DenseNet201} \\ \hline
       Intel i9                     &14.1        &-                    &19.7        &-                      &16.8        &-                 
                                    &20.1        &-                    &19.2        &-                      &13.4        &-                  \\ \hline
       Nvidia TITAN Xp              &4540.0      &-                    &4760.0      &-                      &4234.8      &-
                                    &4925.0      &-                    &4198.6      &-                      &4340.0      &-                  \\ \hline
       Ma, et. al. \cite{16_fix_2}  &-           &-                    &715.9       &0.47                   &611.4       &0.40
                                    &-           &-                    &707.2       &0.47                   &-           &-                  \\ \hline
       RNA\cite{rna}                &687.8       &-                    &878.1       &-                      &804.3       &-
                                    &-           &-                    &-           &-                      &-           &-                  \\ \hline
       {\bf ours}                   &{\bf 1066.4}&{\bf 1.39}           &{\bf 1086.8}&{\bf 1.42}             &{\bf 1101.9}&{\bf 1.43}  
                                    &{\bf 1121.4}&{\bf 1.46}           &{\bf 1121.3}&{\bf 1.46}             &{\bf 1104.7}&{\bf 1.44}         \\ \hline                                
  \end{tabular}
  \label{table:throughput}
\end{table*}

\begin{table*}
  \centering
  \caption{Comparison with prior accelerators on VGG16. "-" means no reported results.}
  \begin{tabular}{c|c|c|c|c|c|c|c}  \hline
     \multirow{2}{*}{}            &Mei, et.al. &Xiao, et. al.  &Ma, et. al.    &Angle-Eye       &RNA        &BFP            &\multirow{2}{*}{{\bf ours}}\\
                                  &\cite{16_fp}&\cite{16_fix_1}&\cite{16_fix_2}&\cite{angle_eye}&\cite{rna} &\cite{bfp_fpga}                            \\ \hline
     Year                         &2017        &2017           &2018           &2018            &2018       &2019           &{\bf 2019}                 \\ \hline
     Platform                     &XC7VX690T   &XC7Z045        &Arria 10 GX1150&XC7Z020         &XC7Z045    &XC7VX690T      &{\bf XC7K325T}             \\ \hline
     Frequency (MHz)              &200         &100            &200            &214             &-          &200            &{\bf 200}                  \\ \hline
     Quantization                 &16-bit      &16-bit         &16-bit         &8-bit           &8/4-bit    &8-bit          &{\bf 8-bit}                \\ 
     Strategy                     &floating    &fixed          &fixed          &fixed           &fixed/log  &block floating &{\bf floating}             \\ \hline
     Top-1/Top-5                  &\multirow{2}{*}{70.46/89.77}&\multirow{2}{*}{-/-}            &\multirow{2}{*}{-/-}       &\multirow{2}{*}{67.72/88.06}               
                                  &\multirow{2}{*}{70.19/89.81}&\multirow{2}{*}{68.31/-}        &\multirow{2}{*}{{\bf 70.05/89.68}}                     \\ 
     Accuracy (\%)                &            &               &               &                &           &               &                           \\ \hline
     DSP Used                     &1728        &824            &1518           &780             &-          &1027           &{\bf 768}                  \\ \hline
     Throughput (GOPS)            &202.42      &229.55         &715.9          &84.3 (CONV)     &878.11     &760.83         &{\bf 1086.8}               \\ \hline
     per DSP Throughput           &\multirow{2}{*}{0.117}      &\multirow{2}{*}{0.279}          &\multirow{2}{*}{0.472}  
                                  &\multirow{2}{*}{0.444}      &\multirow{2}{*}{-}              &\multirow{2}{*}{0.741}     &\multirow{2}{*}{{\bf 1.42}}\\ 
     (GOPS/DSP)                   &            &               &               &                &           &               &                           \\ \hline
     Power (W)                    &10.81       &9.4            &-              &3.5             &7.2        &9.18           &{\bf 9.42}                 \\ \hline
     Power Efficiency             &\multirow{2}{*}{18.72}      &\multirow{2}{*}{24.42}          &\multirow{2}{*}{-}
                                  &\multirow{2}{*}{24.1}       &\multirow{2}{*}{72.8}           &\multirow{2}{*}{82.88}     &\multirow{2}{*}{{\bf 115.4}}\\
     (GOPS/W)                     &            &               &               &                &           &               &                           \\ \hline
  \end{tabular}
  \label{table:vgg}
\end{table*}

\subsubsection{Throughput and per DSP throughput for Different CNNs}
\label{subsubsection:throughput_per_dsp}

Six representative CNNs, including {\it slim, medium} and {\it deep} networks (see Table~\ref{table:benchmark}), are mapped on our processor. When calculating the CNN size, one MAC is counted as two operations. The throughput is measured in GOPS (Giga Operations Per Second), and is reported for different networks on our processor, Intel i9 and Nvidia TITAN Xp in Table~\ref{table:throughput}. For the evaluation on Intel i9 and Nvidia TITAN Xp, we run the CNNs using the Darknet framework with $batch\_size=1$ and 32-bit floating-point data format. As the existing studies \cite{16_fix_2}\cite{rna} support multiple networks, we also include their results in Table~\ref{table:throughput}.

Compared with the existing accelerators, our processor outperforms them in both throughput and per DSP throughput. Particularly, the average improvement of throughput is 63.5\% and 38.6\% compared with \cite{16_fix_2} and \cite{rna}, respectively. Moreover, the average improvement of per DSP throughput is 2.2$\times$ compared with \cite{16_fix_2}. In the approach proposed in \cite{rna}, they use LUT to implement multipliers, so we do not compare per DSP throughput with them. Our FPGA processor outperforms Intel i9 by 64.5$\times$ in terms of throughput because of the high parallelism in our processor. Although Nvidia TITAN Xp can achieve higher throughput (due to more hardware resources) than our processor does, the average power of Nvidia TITAN Xp is 286W, and the average power efficiency of our processor is 6.9$\times$ of the Nvidia TITAN Xp. 

\subsubsection{Comparison with Previous Accelerators on VGG16}
\label{subsubsection:comp_vgg}

We run the classification network VGG16 on our processor, and compare the results with six typical studies, as shown in Table~\ref{table:vgg}. We also list the detailed implementation information, such as platform, working frequency and quantization strategy in Table~\ref{table:vgg}. First, our processor, which uses the LPFP quantization scheme, has a negligible top-1 and top-5 accuracy degradation of 0.33\% and 0.13\%, respectively. Although the work in \cite{16_fp} and \cite{rna} can maintain lower accuracy loss than ours, the approach in \cite{16_fp} uses 16-bit floating-point data format, which results in higher bandwidth and memory requirement and lower per DSP throughput, while the approach in \cite{rna} needs 144 extra hours for the fine-tuning process. Second, our processor outperforms all the six accelerators in terms of throughput and per DSP throughput. Particularly, the improvements of throughput and per DSP throughput are from 24\% to 11.89$\times$ and from 92\% to 11.14$\times$, respectively. These improvements mainly come from the parallel computation pattern in FPFU and the implementation of four 4-bit MACs within one DSP slice. To the best of our knowledge, this is the first work that can simplify the multiplication to 4-bit and implement four MACs inside one DSP slice while maintaining comparable top-1/top-5 accuracy without any re-training process. Finally, we also show the power efficiency in Table~\ref{table:vgg}, and our processor improves the power efficiency by 39\% -- 5.16$\times$.

\subsubsection{Comparison with Previous Accelerators on YOLO}
\label{subsubsection:comp_yolo}

We further compare the detection network YOLO \cite{obj_detec, yolov2} with prior accelerators \cite{16fix_tinyyolo, aristotle_hotchips, angle_eye, fix_tinyyolov2} and we use the tiny version of the YOLO network. The comparison results are shown in Table~\ref{table:yolo}, where we also list the mean average precision (mAP) loss of our quantized networks. Compared with the full precision network, the mAP loss of quantized tiny-yolo and tiny-yolo-v2 is 0.3\% and 0.1\%, respectively. The hardware comparison with prior accelerators shows that our processor is 20.1$\times$ and 49.7$\times$ higher in terms of throughput for tiny-yolo and tiny-yolo-v2, respectively. Moreover, due to the implementation of four 4-bit MACs within one DSP slice, the per DSP throughput improves by 5$\times$ compared with prior accelerators on average.

\begin{table*}
  \centering
  \caption{Comparison with prior accelerators on YOLO. "-" means no reported results.}
  \begin{tabular}{c|c|c|c|c|c|c}  \hline
                                     &Ma, et.al.\cite{16fix_tinyyolo} &Aristotle\cite{aristotle_hotchips} &Angle-Eye\cite{angle_eye} &Wai, et.al.\cite{fix_tinyyolov2}
                                     &\multicolumn{2}{|c}{{\bf ours}}                                                                                  \\ \hline
     Year                            &2017               &2017            &2018              &2018          &\multicolumn{2}{|c}{{\bf 2019}}           \\ \hline
     Platform                        &XC7V485T           &XC7020          &XC7Z020           &Cyclone V     &\multicolumn{2}{|c}{{\bf XC7K325T}}       \\ \hline
     Frequency (MHz)                 &143                &214             &-                 &117           &\multicolumn{2}{|c}{{\bf 200}}            \\ \hline
     Quantization Strategy           &16-bit fixed       &8-bit fixed     &8-bit fixed       &8-bit fixed   &\multicolumn{2}{|c}{{\bf 8-bit floating}} \\ \hline
     Network                         &tiny-yolo          &tiny-yolo       &tiny-yolo         &tiny-yolo-v2  &{\bf tiny-yolo} &{\bf tiny-yolo-v2}       \\ \hline
     mAP loss (\%)                   &-                  &-               &-                 &-             &{\bf 0.3}       &{\bf 0.1}                \\ \hline
     DSP Used                        &112                &198             &-                 &122           &\multicolumn{2}{|c}{{\bf 768}}            \\ \hline
     Throughput (GOPS)               &48                 &36.5            &62.9              &21.6          &{\bf 987.2}     &{\bf 1095.4}             \\ \hline
     per DSP Throughput (GOPS/DSP)   &0.429              &0.184           &-                 &0.177         &{\bf 1.29}      &{\bf 1.43}               \\ \hline
  \end{tabular}
  \label{table:yolo}
\end{table*}

\section{Related Work}
\label{section:related_wrok}

{\bf Weight and Computation Reduction.} CNNs are typically over-parameterized, and extensive accelerator developers in recent years focus on using CNN approximation algorithms, including weight reduction, computation complexity reduction and quantization to accelerate CNN inference \cite{approx_survey}. The accelerator proposed in \cite{wino_fccm17, wino_fpga18} used Winograd algorithm to reduce the number of multiplication in convolution, thus reducing computation complexity. EIE \cite{EIE}, Cambricon-X \cite{cambricon_x} and Cambricon-S \cite{cambricon_s} were the mainstreaming accelerators that benefit from weight and computation complexity reduction techniques. However, the irregularity caused by these algorithms degrades the parallelism and hardware efficiency \cite{clstm}.

{\bf Quantization.} Accelerators with quantization is another concentration. XNOR\_Net \cite{xnor_net} applied weights binarization by quantizing weights into \{-1, 1\} with a scaling factor for AlexNet. The lightweight YOLOv2 \cite{bin_yolo} was another binarization approach which focused on object detection CNN. Accelerator with ternary representation, which added zero to the binary set, was introduced to help improve the accuracy \cite{tnn}. Although these accelerators achieve remarkable power and storage saving, they both suffer from significant accuracy loss. Moreover, they all need time-consuming re-training process to compensate for the quantization error. 16-bit quantization oriented accelerators, including floating-point and fixed-point representations, solved the problem of accuracy loss \cite{16_fp, 16_fix_1, 16_fix_2, 16fix_tinyyolo}. However, the storage requirement is still huge, and the per DSP throughput is extremely low (less than 0.5GOPS/DSP) because of the usage of 16-bit. 

8-bit quantization makes a trade-off between storage and accuracy. The accelerators \cite{angle_eye, rna} optimized the computation patterns with 8-bit fixed-point quantization to improve the performance for different CNNs. DNNBuilder \cite{dnnbuilder} was proposed to automatically build DNN accelerators to satisfy the performance and power efficiency demands on embedded and cloud FPGAs, while Cloud-DNN \cite{cloud_dnn} was the framework for mapping DNN models to cloud FPGAs. Block floating-point scheme with 8-bit mantissa was used in \cite{bfp_fpga} to accelerate the inference of CNN while maintaining accuracy. However, all these accelerators need 8-bit MAC to perform convolution, leading to a low per DSP throughput (less than 0.8GOPS/DSP). A more aggressive method quantized the small values of the weights into 4 bits and keeps the remaining 16 bits as full precision, by dividing the weights into the low-precision and high-precision regions according to the values of the weights \cite{OLAccel}. HAQ \cite{haq} proposed a mixed precision quantization approach with a trade-off between quantization policy and hardware performance. However, both studies need time-consuming re-training process to compensate for quantization errors.

Different from all the above methods, the proposed LPFP quantization scheme fully exploits the properties of weights and activations, thus obtaining a comparable or better accuracy for {\it deep} CNNs. Moreover, the LPFP quantization method gets rid of the time-consuming re-training process that needs labelled data and extra computing, because access to labelled data can be difficult in practice as hardware and CNN algorithms are often developed by different parties. Furthermore, with the help of the LPFP quantization method, our processor only needs 4-bit MACs, thus dramatically improving the per DSP throughput. Overall, the proposed processor achieves better performance on FPGA.

\section{Conclusion}
\label{section:conclusion}

We have proposed a low precision floating-point quantization method, called LPFP, to reduce memory size and memory access with negligible accuracy degradation (less than 0.5\% for top-1 and 0.3\% for top-5 accuracy) for CNN interference. LPFP does not need any re-training. Furthermore, we have reduced the bit width for multiplication to 4-bit with comparable accuracy and implemented four 4-bit MACs within one DSP slice in Xilinx Kintex 7 FPGA family. Experiments using Xilinx KC705 platform and six typical CNN networks show that we achieve an average throughput and per DSP throughput of 1100.4 GOPS and 1.43 GOPS, respectively. Moreover, the average throughput is 64.5$\times$ and 1.5$\times$ over Intel i9 and existing accelerators, respectively. Particularly for VGG16 and YOLO, we outperform six existing accelerators in terms of average throughput by 3.5$\times$ and 27.5$\times$, while improving per DSP throughput by 4.1$\times$ and 5$\times$, respectively.
To the best of our knowledge, this is the first in-depth work that can simplify the multiplication to 4-bit and accommodate four MACs in one DSP slice while maintaining comparable top-1/top-5 accuracy without any re-training.


%

\ifCLASSOPTIONcaptionsoff
  \newpage
\fi



\bibliographystyle{IEEEtran}
\bibliography{ref}
\end{document}